\documentclass[twocolumn,superscriptaddress,floatfix,pre,showpacs]{revtex4-1}
\usepackage{graphicx}
\usepackage{amssymb,amsfonts,amsmath,fancybox,epstopdf,epsfig,color, graphicx}

\begin{document}

\title{Resilience and rewiring of the passenger airline networks in the United States}

\author{Daniel R.
Wuellner}
\email{dwuellner@math.ucdavis.edu}
\affiliation{Graduate Group in Applied Mathematics, University of California, Davis, CA 95616}

 \author{Soumen Roy}
 \email{sroy@cse.ucdavis.edu} 
 \affiliation{Department of Mechanical and Aeronautical Engineering, University of California, Davis, CA 95616}
 \affiliation{Department of Medicine and Institute for Genomics and Systems Biology, The University of Chicago, Chicago, IL 60637, USA}
 
 \author{Raissa M. D'Souza}\email{raissa@cse.ucdavis.edu}
 \affiliation{Graduate Group in Applied Mathematics, University of California, Davis, CA 95616} 
 \affiliation{Department of Mechanical and Aeronautical Engineering, University of California, Davis, CA 95616} 
 \affiliation{Department of Computer Science, University of California, Davis, CA 95616} 
 \affiliation{Santa Fe Institute, 1399 Hyde Park Road, Santa Fe, New Mexico, USA}

\begin{abstract}
The air transportation network, a fundamental component of critical infrastructure, is formed from a collection of individual air carriers, each one with a methodically designed and engineered network structure. We analyze the individual structures of the seven largest passenger carriers in the USA and find that networks with dense interconnectivity, as quantified by large $k$-cores for high values of $k$, are extremely resilient to both targeted removal of airports (nodes) and random removal of flight paths paths (edges). Such networks stay connected and incur minimal increase in an heuristic travel time despite removal of a majority of nodes or edges. Similar results are obtained for targeted removal based on either node degree or centrality. We introduce network rewiring schemes that boost resilience to different levels of  perturbation while preserving total number of flight and gate requirements.  Recent studies have focused on the asymptotic optimality of hub-and-spoke spatial networks under normal operating conditions, yet our results indicate that point-to-point architectures can be much more resilient to perturbations. 
\end{abstract}
\pacs{89.40.Dd, 89.75.Hc, 89.75.Fb}


\maketitle

Air travel is a principal means of fast and effective transportation of people and goods over large distances across countries or continents, around the globe. It is critical to the  functioning of countries and the world economy as a whole. The aggregate network of air travel worldwide built by considering all flights amongst all destinations throughout the globe (the world airline network) has been the subject of much recent study~\cite{AmaralPNAS,GuimEPJB04,BarratPNAS04,ColizPNAS06,GuimPNAS05}.  The focus has been on analysis of overall flow patterns and the consequences for the spread of global epidemics~\cite{ColizPNAS06}, as well as identifying the overall importance of individual airports~\cite{GuimPNAS05}.  An aggregate level analysis has also been carried out on the airline networks of a few individual countries by studying their temporal evolution\cite{brazilJSM09} or by uncovering similarities  with the world airline network, namely ``scale-free" and small-world characteristics~\cite{airChinaPRE04,airIndiaPhysica08}. 

Our interest is not in overall flow, but in design and operation of critical infrastructure. The aggregate view of air travel is built up from a collection of co-existing airline networks, operated independently by distinct entities. Each independent operator must build a well-connected and economically successful airline network which is resilient to random or systematic vagaries, 
ranging from acts of nature to terrorism. Furthermore, an individual airline has direct control only over its own network, thus understanding changes to an individual network structure  that can lead to improved efficiency and resilience are quite relevant. 

Herein we analyze and contrast the network structures of the seven largest passenger airlines in the United States of America (USA). Small-world attributes are exhibited by the network of each carrier, yet, rather than scale-free power law distributions, we find that the distribution in airport connectivity is better described by either a simple exponential decay or a cumulative log-normal distribution. More pronounced than distribution in connectivity, we find that Southwest Airlines (SW) stands apart from the other six carriers by its k-core structure (defined in detail below) and its extreme resilience to random or targeted deletion of nodes (airports) or edges (flight paths).  Edge deletion corresponds to, for instance,  weather preventing travel between two airports, while node deletion corresponds to closure of an airport.   SW has essentially built a core network, comprising more than half of its overall destinations, which is a dense mesh of interconnected high-degree (i.e., ``hub") airports.  We explore the interplay between placing hubs in the  periphery versus the core of a network and introduce a general network rewiring process which keeps constant the demand on each node and the amount of flow between nodes, that enhances the k-core structure and increases resilience of a network.   

One fundamental consideration when building a new airline network, or expanding an existing one, is whether to prefer ``point to point" (PP) or ``hub and spoke" (HS) connectivity.  
In the PP scenario, a passenger can travel on a direct non-stop flight to a range of destinations at shorter distances, but to travel considerable lengths has to transit and take multiple flights.  In the HS scenario, in contrast, a passenger can travel non-stop only to a few central hubs, and from there transit to their final destination (almost always requiring two-hops unless the hub is their ultimate destination).  Rigorous analysis shows {\em asymptotic} optimality of HS models for spatial transportation networks with transfer costs~\cite{aldous}. Analytic arguments, backed by numerical simulations 
indicate that HS architectures are optimal for travelers wishing to minimize the number of connecting flights required instead of overall distance travelled~\cite{newman-gastner}. 
Inspired in part by studies on airport networks,  a  general model of weighted networks via an optimization principle was proposed 
 in which a clear spatial hierarchical organization, with local hubs distributing traffic in smaller regions, emerges as a result of the optimization~\cite{barthelemy}.   Thus there seems to be a growing consensus in the literature regarding HS structures arising out of optimization of resources. However, real-world structures need to also be resilient and robust.  As show herein, PP structures can be much more resilient than HS structures. 
 
The majority of the larger airlines operating in the USA at present predominantly follow the HS pattern. This was not the case prior to 1978, when the USA Federal Government regulated air traffic, with special attention paid to ensure lower traffic  routes were not ignored~\cite{deregulation}, effectively enforcing PP architectures.  Once deregulated in 1978, most airlines gradually shifted to their current HS pattern. A significant exception was Southwest Airlines (SW), which continued to build a PP system. 

\begin{figure}[t!b]
\includegraphics[width=0.75\columnwidth,angle=270]{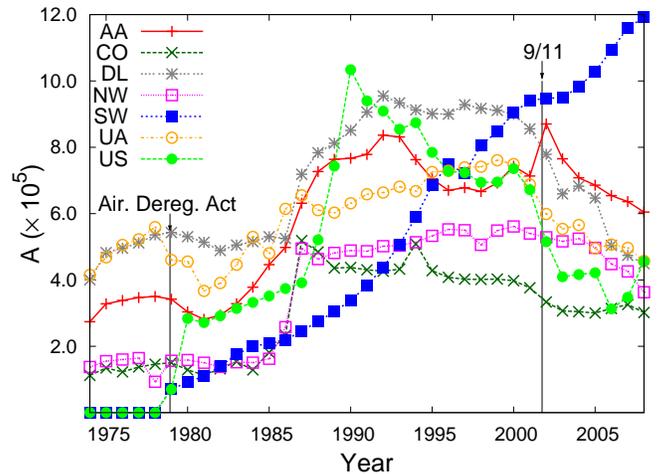}%
\caption{(Color online) Annual domestic departures (A) for the major U.S. airlines for each year between 1974-2008, mined from \cite{dot_long}.}
\label{longitudinal}
\end{figure}

As of the end of 2007  (the focal year for our data collection) SW was the largest airline (by both number of domestic passengers and domestic departures) not only in the United States, but also in the entire world~\cite{WATS}.  Its sheer size together with the extremely consistent economic success of SW~\cite{SW-profit}  provide strong evidence for the efficacy of PP networks.  As shown in Figure~\ref{longitudinal}, while the major carriers experienced dramatic growth after deregulation, all except SW stagnated by 1992.  SW continued to grow throughout the entire period, and surpassed all of the carriers in terms of annual departures by 2000.  Throughout its growth, starting from a handful of airports to its current size, the SW network has maintained a PP structure.  It is notable that SW is the smallest carrier by the number of airports served, but the airports that it does serve are on average larger than those served by the other carriers (except US Airways), with an average $6 \times 10^6$ passengers leaving an airport served by SW during the 2007 calendar year.
Ryanair and Easyjet are two examples of successful PP carriers in Europe~\cite{ryanair}. Innovative management policies have also played an important part in the success of SW and are studied extensively in business literature (for instance, Ref.~\cite{sw:management}).

%
\begin{table*}[tbh]
\centering
\caption{Basic network properties of the carriers. $N$ and $E$  denote the number of nodes and edges respectively, and $\left<q\right>$ the mean node degree. $\left<l \right>$ and $\left<C \right>$ denote respectively the mean of the geodesic and clustering coefficient distributions. $r$ and $G(q)$ denote degree assortativity and Gini coefficients. $\alpha(q)$ is the skewness of the degree distribution.}
\begin{tabular*}{\hsize}{@{\extracolsep{\fill}}|l|r|r|r|c|r|c|c|r|}
\hline 
Carrier  & N \ & E \  & \ $\left<q\right>$ \ & $\left<l\right>$ & $\left<C\right>$ & $r$ & $G(q)$ & $\alpha (q)$ \\
\hline
\ & \ & \ & \ & \ & \ &  \ & \ & \\
SW & 64 & 892 & 27.88 & 1.542 &  0.731  &-0.177  & 0.254 & 226.3\\
US & 96 & 556 & 11.58 & 1.990 &  0.672  &-0.367  & 0.521 & 1053.8\\
CO & 117 & 736 & 12.58 & 1.935 &  0.628  &-0.330  & 0.512 & 1742.8\\
UA & 121 & 737 & 12.18 & 1.983 &  0.640  &-0.320 & 0.498 & 1839.6\\
AA & 121 & 1163 & 19.22 & 1.889 &  0.646  &-0.280 & 0.461 & 1542.0\\
NW & 132 & 753 & 11.41 & 2.023 &  0.624  &-0.269 & 0.493 & 2130.1\\
DL & 133 & 906 & 13.62 & 1.943 &  0.586  &-0.272 & 0.499 & 2168.7\\
NW+DL & 163 & 1529 & 18.76 & 1.985 &  0.617  & -0.256 & 0.497 & 2682.7\\
UPS  & 107 & 606 & 11.33 & 1.929 &  0.620  &-0.249 & 0.427 & 1618.7\\
FX  & 334 & 1355 & 8.11 & 3.060 &  0.579  &-0.047 & 0.548 & 1457.1\\
Agg7 & 197 & 3505 & 35.58 & 1.926 &  0.710  &-0.244 & 0.497 & 2993.1\\
AggPass&817  &   9688  &  23.72 & 3.181&  0.639 & 0.185& 0.630& 8758.7\\
AggAll& 1258 & 17437 & 27.72 & 3.005 &  0.557 & 0.097 &0.677 & 17484.5\\
\ & \ & \ & \ & \ & \ & \ & \ &   \\
\hline

\end{tabular*}
\label{airnets-table}
\end{table*}

Here our focus is on network infrastructure with a view to efficiently design or restructure individual networks so they are well-connected, robust and resilient to disturbances.  These findings provide theoretical insight and may be relevant to entities engaged in designing or altering large-scale airline networks, for instance, operators expanding airline networks in developing nations, carriers needing to {\it shrink} an airline (i.e., eliminate flights with minimal impact), and carriers needing to assess the quality of network infrastructure which would result from a merger with another carrier.

\section{The networks} 
All certificated USA air carriers are required to file monthly reports with the USA Department of Transportation, Bureau of Transportation Statistics, detailing information on every flight segment flown during that month.  This information is maintained in a public database~\cite{dot}, from which we download information on every ``scheduled passenger service" class flight segment flown by each of the seven largest U.S. passenger carriers for the entire 2007 calendar year. To isolate the structure of passenger carriers we neglect the small fraction of flights by these carriers which are designated by the ``cargo" (only) class or ``non-scheduled passenger service" (charter) class. Yet, in order to compare the structure of a passenger carrier with a cargo-only air carrier, we also download all flights flown during the 2007 calendar year by two cargo-only carriers  (Federal Express and United Parcel Service). We neglect scheduling and restrict ourselves to the domestic routes of international carriers.

The seven largest US passenger airlines (by number of passengers flown) are in order, Southwest (SW), American Airlines (AA), Delta (DL), United Airlines (UA), Northwest (NW), US Airways (US), and Continental (CO).  These seven carriers account for 61.6\% of all domestic passengers enplaned in 2007. For each carrier $c$ we construct two distinct views of the network. The first, denoted $G^c(N^c,E^c)$, is a binary view capturing connectivity (i.e., which airports are connected via direct flights). The second, denoted $W^c(N^c,E^c)$,  captures both connectivity and the total number of flights flow between airports.  
To explicitly construct $W^c(N^c,E^c)$ a directed edge is added from each origin airport to its destination airport, with edge weight equal to the total number of flight segments from that origin to that destination flown by carrier $c$ in 2007.  The unweighted (binary) version of this graph is $G^c(N^c,E^c)$, and is the equivalent of the ``route map" for that carrier. 
The vertices in both views, $N^c$,  are the set of all airports listed as an origin or destination airport for carrier $c$ which are also included in that carrier's list of official domestic destinations as stated on June 2008.  This data ``scrubbing'' step eliminates airports used only for diverted aircraft (which have substantially fewer numbers of flights than official airports and otherwise introduce noise).  

We consider both node degree and strength. The out-degree of node $i$, $q^{out}_i$,  is  the 
number of distinct destinations that can be reached directly from $i$. The in-degree, $q^{in}_i$ is the number of distinct incoming origins.  We find $q^{in}_{i} \sim q^{out}_{i}$ (airports are almost always connected in both directions) so simply denote node degree as $q_i$. We also consider  the ``strength", $s_i$ of the $i$'th node,  defined as in Ref~\cite{BarratPNAS04}. The in-strength (out-strength) of an airport is the total number of flights landing (departing) there, for that specific carrier, in 2007. Formally, the in-strength (out-strength) is the sum over all edge weights in $W^c(N^c,E^c)$ for edges terminating (originating) at that node.  We find $s^{in}_{i} \sim s^ {out}_{i}$; so for the remainder we treat all edges as undirected and set the undirected edge weights in $W^c$ to be the maximum edge weight in either direction.

In addition to the networks of individual carriers, we construct three different views of the aggregate airline network of the USA: Agg7, which is the aggregate over the seven largest passenger carriers; AggPass, which is the aggregate over all ``scheduled passenger service" class flights flown during 2007 by all carriers (not just the seven largest); finally, AggAll is the aggregate over every single flight segment flown during 2007, regardless of service class or carrier.  Formally, to construct the distinct aggregate views we take the union over all nodes and edges for the set of carriers involved: $G^{Agg}(N^{Agg},E^{Agg})$ where $N^{Agg}= \bigcup_c N^c$ and $E^{Agg}=\bigcup_c E^c$ and $W^{Agg}(N^{Agg},E^{Agg})$, where $E^{Agg}$ is the sum over all the corresponding edge weights.  Finally, in light of a merger between two carriers  (NW and DL) which took place in early 2008~\cite{merger} we construct their merged networks,  $G^{NW+DL}$ and $W^{NW+DL}$.

\section{Characterization}
\subsection{General metrics}

We first compare the network structures of the distinct airlines.  Results are summarized in Table~\ref{airnets-table}, with the passenger airlines listed in order of increasing number of airports serviced ($N$). Also included are the results for the three different aggregate views (Agg7, AggPass, and AggAll), the two cargo carriers Federal Express (FX) and United Parcel Service (UPS), and the ``NW+DL" network. The number of distinct direct connections between airports for each carrier is listed as $E$ (the total number of edges in $G^c(N^c,E^c)$).  The average airport degree for each airline network, denoted $\left<q\right>$, is simply $\left<q\right>=2E/N$.  The average shortest path length over all source-destination pairs is denoted $\left<l \right>$. (This is the average number of flight segments required to fly from any airport in the network to any other.)  The average clustering coefficient~\cite{WSNature98} is denoted $\left<C \right>$.

For comparison, we generate a corresponding Erd\H{o}s-R\'{e}nyi (ER) random graph for each carrier, using that carrier's $N$ and $E$ values.  The values of $\left <l\right>$ and the average value of betweeness centrality~\cite{betweenness}  for the actual carriers agree almost exactly with the values for the corresponding ER realizations, strongly suggesting that density alone determines these two properties.  All remaining properties show significant differences between the real networks and ER equivalents. Note, all carriers have $\left<l \right> < \ln N$ and values of $\left<C \right> > \left<C_{ER} \right>$, thus can be considered ``small-world" networks.  It is noteworthy that SW has $\left <l\right>\approx 1.5$, with the remaining carriers all having $\left <l\right>\approx 2$ (requiring two-hops between most source-destination pairs).

\begin{figure*}[bth]
\includegraphics[width=1.025\textwidth]{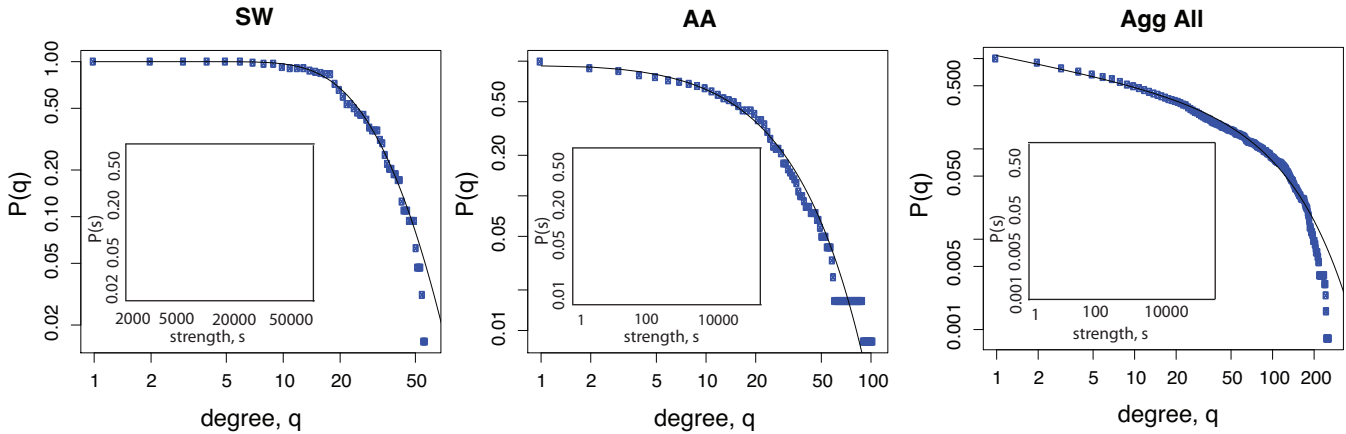}
\caption{(Color online) Cumulative degree distribution ($P(q)$) with cumulative strength distribution ($P(s)$) inset for (a) SW, (b) AA (which is representative of the other carriers), and (c) the aggregate over all flights flown in 2007.  The points indicate the empirical data. The solid lines are the best fit theoretical distribution where appropriate.  For SW both $P(q)$ and $P(s)$ are best fit by a cumulative log-normal.  $P(q)$ for AA is best described by an exponential. For the aggregate over all flights both $P(q)$ and $P(s)$ are well described by a power-law with exponential decay.}
\label{ddist}
\end{figure*}

To quantify the extent to which a network follows the ``hub and spoke" (HS) pattern, the degree assortativity coefficient~\cite{MEJN-mixing}, $r$,  seems a natural choice. 
$r>0$ indicates a tendency of high-degree nodes to connect to other high-degree nodes. $r<0$ indicates a tendency of high-degree nodes to connect to low-degree nodes. Thus, a larger negative (dissassortative) value of $r$ should indicate that the network follows the HS paradigm more closely. Previous studies have found the airport networks of China and India and the airline networks of  European carriers to be strongly disassortative (Refs.~\cite{airChinaPRE04,airIndiaPhysica08,europe} respectively), while in contrast the world airline network shows assortative behavior~\cite{BarratPNAS04}.  

As can be seen in Table~\ref{airnets-table}, we find that all the individual carriers as well as their aggregate view (Agg7) have dissassortative structures, yet AggPass is assortative, and FX and AggAll have values of $r$ close to zero. The value of $r$ for SW is about half the magnitude of the other passenger carriers as would be expected given SW's predominantly PP structure. However the value of $r$ for FX is significantly smaller in magnitude than that for SW, though we explicitly observe that the topology of FX exhibits strong HS structure.  In this context,  we turn to a measure used in the transportation literature~\cite{gini:transport} to quantify the extent of HS structure, the Gini coefficient~\cite{sen}. 
The degree Gini coefficient, $G(q)$, is defined for a network of size $N$ as, 
\begin{equation}
G(q)=\frac { \sum_{i=1}^{N} \sum_{j=1}^{N} |q_i-q_j| } {N^2 \left<q\right>}, 
\end{equation}
where $\left<q\right>={2}E/N$. It essentially measures the magnitude of the difference in node degree between all pairs of nodes in a network normalized by average node degree.  
As seen in Table~\ref{airnets-table}, the Gini coefficient metric correctly indicates the HS structure of FX. 
Likewise, the values of $G(q)$ indicate extremely strong HS structures for AggPass and AggAll, while the values of $r$ indicate assortative, PP structures. 
The Gini coefficient has been widely used in fields such as economics~\cite{sen} and ecology~\cite{ecology}. Our findings indicate the Gini coefficient more accurately captures the HS versus PP nature of a network than does the assortativity coefficient. 

The assortativity coefficient is by definition a correlation coefficient and it is a well known that correlation coefficients are extremely sensitive to outliers~\cite{Outliers}. Federal Express officially reports that their network has a ``superhub" in Memphis, Tennessee (which also ranks as the world's largest cargo airport)~\cite{Fedex}. Memphis thus acts as an outlier and changes the value of assortativity that would otherwise have been expected for FX. The vast majority of commercial carriers have a HS structure and when we merge all the networks together to create the AggPass and AggAll views, a few superhubs may arise as an artifact of merging the common hubs of many carriers.  
This appears to be the cause of the positive values of assortativity for AggPass and AggAll (where large values of the Gini coefficient in both these cases would lead us to expect disssortative networks). Notably, in Agg7, such an unexpected value of assortativity is not witnessed (which in part is due to the PP structure of SW which counteracts to some extent the HS structure of other six passenger carriers).

We carried out a detailed analysis of betweenness centrality~\cite{betweenness} in the manner of Ref.~\cite{GuimPNAS05}, for all the passenger airlines.  For a few airlines, we do find examples of airports with betweeness values that are relatively higher than their degree (e.g., IAH for CO, PHX for US, STL for AA and LAX for DL). However, this mismatch is not as strongly disproportionate as that of say the Anchorage airport in Ref.~\cite{GuimPNAS05}. Hence, we classify our observation as ``weak anomalous centrality".

We analyze the distribution of node degree and node strength, with $p(q)$ the observed probability of a carrier having a node of degree $q$ and $p(s)$ the observed probability of having a node with strength $s$.  The raw probability distributions are noisy, thus we construct the complementary cumulative distributions $P(x) = \sum_{i \ge x} p(x)$.  These cumulative distributions are right-skewed for each carrier, with the value of degree distribution skewness given in Table~\ref{airnets-table} under $\alpha(q)$. (Note that the skew for SW is an order of magnitude less than that for other carriers.)

We also analyze how well each empirically observed degree distribution and strength distribution can be fit by a theoretical distribution, considering the following forms: 1) power law, 2) exponential, 3) stretched exponential, 4) power law with exponential decay, 5) cumulative log-normal distribution.  We use the nonlinear least squares fitting routine of the R Statistical Computing platform~\cite{R-project} to solve for the parameters values for each candidate distribution which provide the best fit to the data. Finally, we calculate the residual sum of squares between these best fit candidate distributions and the empirical data.  In almost all cases, one of the candidate distributions clearly minimizes this difference. 
Although there exist more rigorous methods for extracting the best fit power law exponent to a data set~\cite{CSN}, the airline networks analyzed herein are too far from power law distributions to warrant the overhead associated with such techniques.

Figure~\ref{ddist} shows the results for SW, for AA (representative of the other carries) and AggAll (the aggregate over all flights flown in 2007).  Focusing on the cumulative degree distribution, $P(q)$, the SW  network is best described by the cumulative log-normal distribution.  The other six individual carriers all have networks with $P(q)$ well described by simple exponential distributions. 
Likewise, the theoretical distribution which best describes the aggregate over the seven passenger carriers (Agg7) is a simple exponential distribution.  The aggregate over all passenger carriers (AggPass) is best described by a cumulative log-normal distribution, while the aggregate over all flights flown in 2007 (AggAll) by a power law with exponential decay. 
Turning to strength distributions, $P(s)$, SW is again best described by a cumulative log-normal, and the aggregate over all flights flown in 2007 is by a power law with exponential tail. Although the distributions are broad, all of the distinct aggregate views have tails decaying more sharply than exponential.

\subsection{k-core structures}
The SW network is distinguished from the networks of the other carriers by the metrics of Table~\ref{airnets-table}, yet the difference in topology is even more pronounced when the $k$-core structures of the distinct carriers are compared.
The $k$-core of the network is a subgraph constructed by iteratively pruning all vertices with degree less than $k$~\cite{kcore83,kcore84}.  For instance, starting from an original network we remove all nodes with degree $q< k$ and their corresponding edges, then successively remove all nodes (along with their edges) which are now of degree $q < k$ in the pruned network, and continue iterating until all remaining nodes have $q \ge k$.  The remaining subgraph is the $k$-core.  We also consider the $k$-shell, which consists of all nodes which are present in the $k$-core but not in the $(k+1)$-core.  Likewise, the ``coreness" of node $i$, denoted $c_i$, is defined as the largest value of $k$ for which the node is a member of the $k$-core.  $k_{\rm max}$  denotes the maximal coreness within a network ({\it i.e.}, the value of the maximum $k$ for which the network has a non-zero $k$-core).

The $k$-core decomposition is a computationally inexpensive way of revealing additional details about the structural role of nodes beyond their degrees and has lately been the focus of several studies in network theory. It  has been used to  predict protein functions from protein-protein interaction networks and amino acid sequences~\cite{kcore-prediction} and to identify the inherent layered structure of the protein interaction network~\cite{almaas}. More recently, the method of $k$-shell decomposition has been used to arrive at a model of internet topology at the autonomous systems level~\cite{havlin-kcore} and to generate random graphs with a specified ``$k$-core fingerprint'' which simulate the autonomous systems network of the internet~\cite{baur-kcore} .

\begin{figure}[tbp]

{\includegraphics[width=0.69\columnwidth,angle=270]{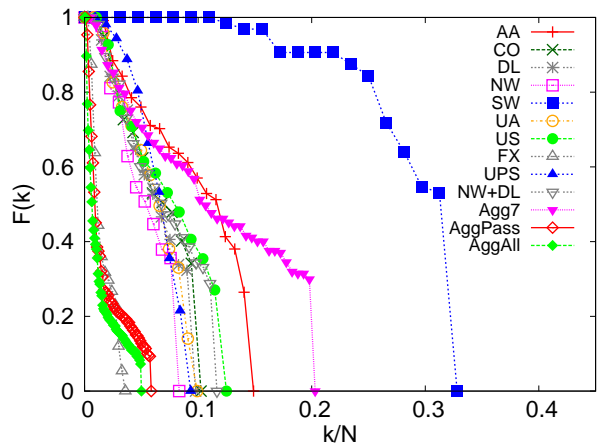}}%
\caption{(Color online) Cumulative $k$-core distribution, $F(k)$, of the largest passenger carrier airline networks, selected cargo carriers, and three different aggregate views.}
\label{kcoreCDF}
\end{figure}

Figure~\ref{kcoreCDF} shows the $k$-core structure of all the carriers studied herein.  Here $F(k)$ is the fraction of all nodes with coreness greater than or equal to $k$.  Note that for SW all nodes $i$ have $c_i \geq 7$, and the majority of nodes have extremely large coreness. 
Two key quantitative differences are prominent when comparing the $k$-core structure of SW to the other carriers: the value of $k_{\rm max}$ and the occupancy of the $k_{\rm max}$ shell.  
For $k_{\rm max}$, in spite of having the smallest number of nodes $N$, SW achieves the highest $k$-core, with value $k_{\rm max}=20$,  and normalized value $k_{\rm max}/N=0.312$.  (The next largest is American Airlines (AA) with $k_{\rm max}=17$ with normalized value $k_{\rm max}/N=0.140$.) With respect to occupancy, that of the largest shell in SW is especially remarkable, with  $53\%$ of all airports belonging to the $k_{\rm max}$-core.  In contrast, for AA,  $26\%$  belong to the $k_{\rm max}$-core.

For all the individual airlines studied here, the highest $k$-shell contains that carrier's hubs and consequently its most viable transfer points. This is consistent with prior work suggesting that the core of a network plays a special role in enhancing navigability of networks where global structural information is unavailable~\cite{BKC-08}.  The large value of $k_{\rm max}$ for SW and the large occupancy of the $k_{\rm max}$-shell suggest that there are many redundant transfer points in the SW network in the cases where a direct connection is not available between source-destination pairs.

\section{Resilience}
\begin{figure*}[htbp]
\boxput*(0.5,0.48) { \includegraphics[width=0.31\columnwidth, angle=270]{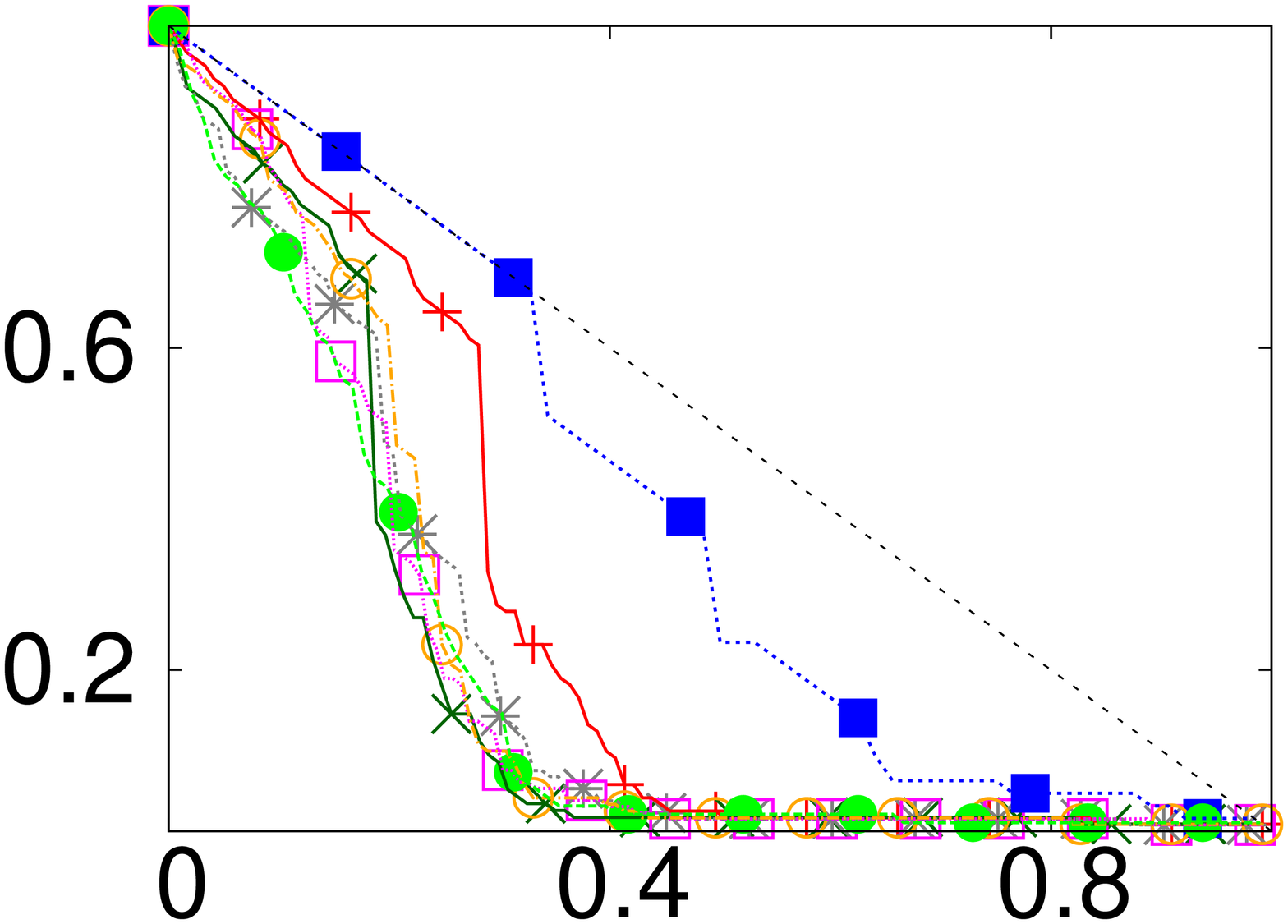} }%
{ \includegraphics[width=0.7\columnwidth,, angle=270]{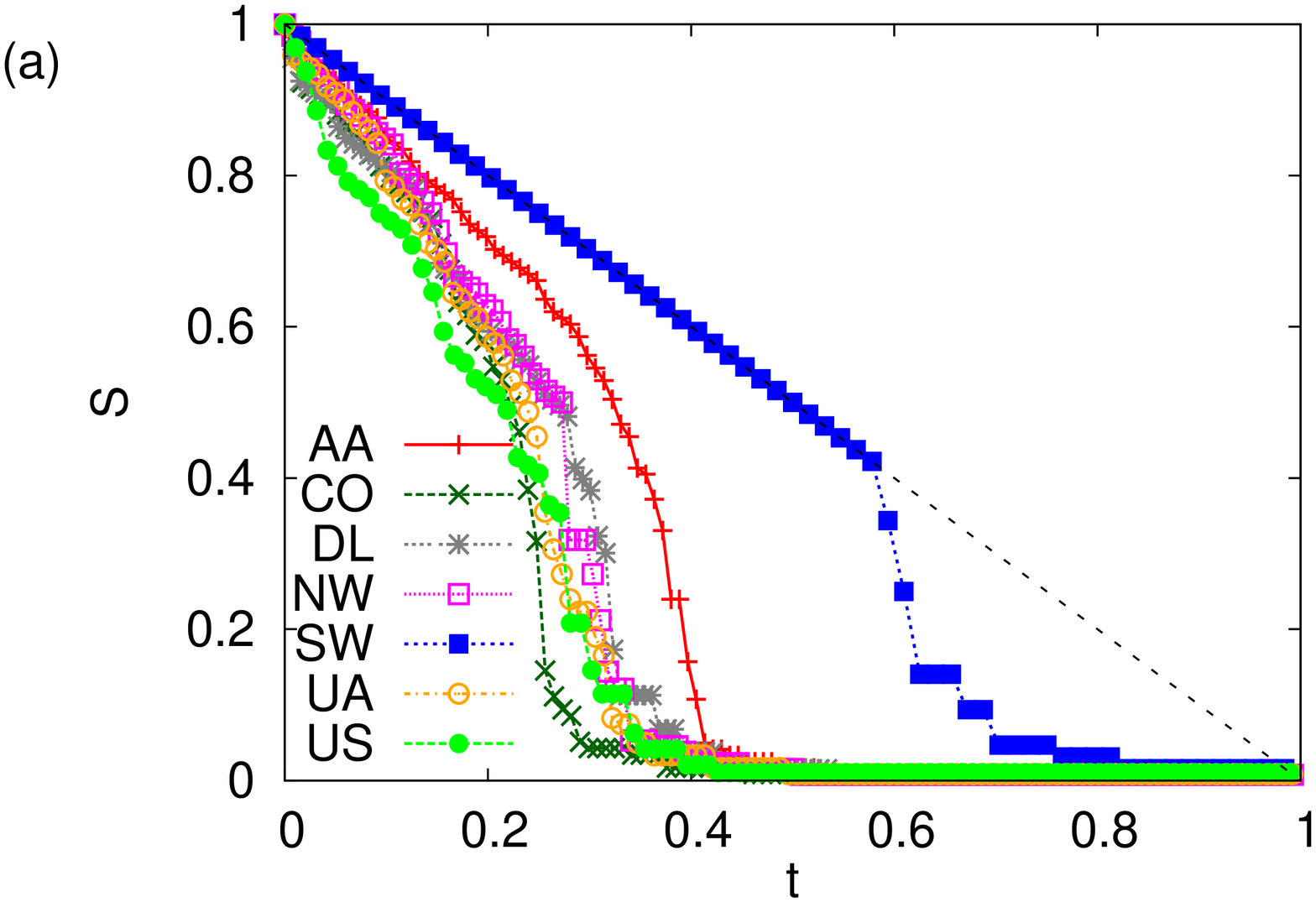}}%
\vspace{0pt}
{ \includegraphics[width=0.7\columnwidth, angle=270]{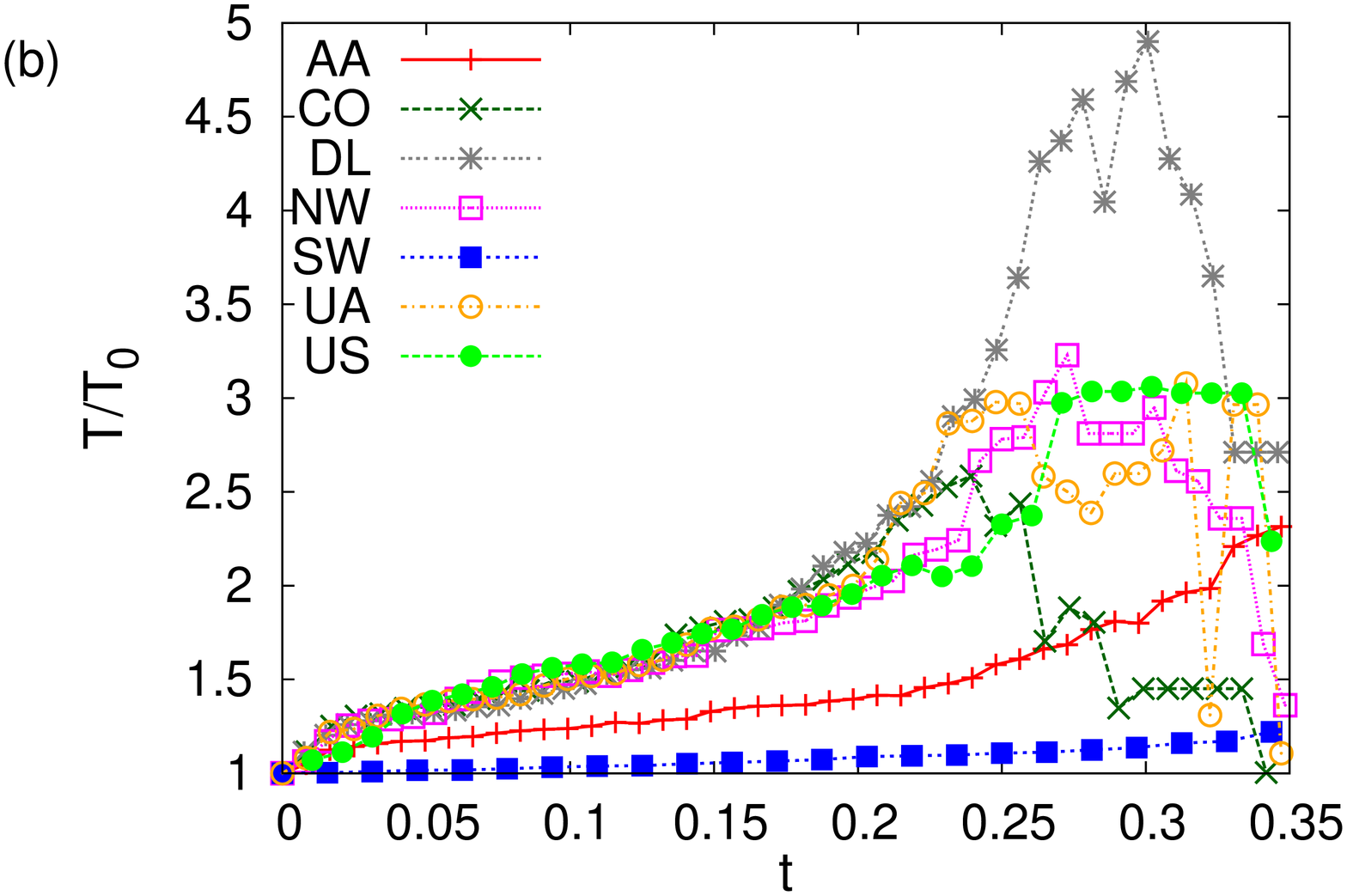} }
\caption{ (Color online) (a) $S$ of each passenger carrier's network as a function of the proportion of nodes removed by degree targeted attack ($t$). 
Targeting by betweenness (inset) rather than degree causes more rapid breakdown of each carrier's network.  The dashed diagonal line depicts the maximal size of $S$ under this process for any network (i.e. the size of $S$ for the corresponding complete graph).  (b)  Normalized travel cost metric $\frac{T(M)}{T_0(M)}$ evaluated on the largest connected component of each passenger carrier's network as a function of $t$.}
\label{figure:removal}
\end{figure*}

We examine the individual passenger carrier's resilience to random edge deletion and targeted and random node deletion. 
Edge deletion corresponds to, for instance, disturbances such as weather preventing travel between a pair of airports (i.e., deletion of a flight path).  Node deletion corresponds to the closure of an airport. 
There is extensive literature investigating various real and simulated networks' resilience to both random and targeted node and edge removal.  One of the first works in this area found that random uncorrelated power-law networks are robust to random node deletion but vulnerable to targeted attack~\cite{AJB2000}.
Different targeted attack strategies have hence been investigated 
using a variety of metrics, notably average inverse geodesic distance (also called `network efficiency') and the relative size of the largest connected component \cite{holmePRE}.  The robustness of graphs with  various kinds of degree distributions have also been studied recently, e.g. in Refs.~\cite{prl:robustness,epl:robustness}  and references therein.

To quantify the performance of the networks under the various deletion processes, we use two topological measures: the size of the largest connected component (denoted $S$) and a relative global travel cost metric (denoted $T$) which accounts for both spatial (geographic) distance and geodesic distance (hop-count).

The metric $T$ is defined by summing over the travel times of the shortest paths through a network.  For a path between $i$ and $j$ consisting of a sequence of edges (denote these $(i,i_1),(i_1,i_2),\ldots,(i_m,j)$), we calculate the total geographic length $d_{ij}$ by adding the length of the edges (geographic length of each edge is available in the U.S. D.O.T. database):
\begin{equation}
d_{ij}=l_{ii_1}+l_{i_{1}i_{2}}+\ldots+l_{i_{m}j}.
\end{equation}
Next we convert the geographic path length to a `flight time' by dividing by a characteristic velocity ($v=804.7$ km/hour = $500$ miles/hour), and for each of the $m$ intermediate nodes in the path we add a fixed `transfer cost' of $\theta=1.0$ hour to account for layover time to give the travel time of the path:
\begin{equation}
t_{ij}=\frac{d_{ij}}{v} + m\theta.
\end{equation}

For each network, we calculate the path with the shortest travel time for every possible source-destination pair $(i,j)$ using Dijkstra's algorithm~\cite{dijkstra}, as implemented in the NetworkX package~\cite{networkx}, by assigning edge weights to each edge $(k,l)$ in $G^c(N^c,E^c)$ corresponding to $d_{kl}+v\theta$.  We must include the transfer cost in each edge to ensure that the shortest path actually minimizes our heuristic flight time and not simply geographic distance.

Finally, we can define the travel cost for the whole network or for just a subset of nodes in the network $M \subseteq N^c$ as the sum over all of the included path costs:
\begin{equation}
T(M)=\frac{1}{2} \sum_{i \in M} \sum_{j \in M} t_{ij}.
\end{equation}
Note, the travel cost over the entire network is $T(N)$. 

Once some nodes are disconnected,  there is no path to any of these disconnected nodes so the travel cost over the whole network is formally infinite.  Consequently, when calculating the travel cost we consider only the nodes in the largest connected component of the randomly damaged graph. We calculate the travel cost between all source-destination pairs in this subset in the original graph, $T_0(M)$, and in the damaged graph, $T(M)$ to obtain the relative travel cost of the damaged network $\overline{T}=\frac{T(M)}{T_0(M)}$. In this manner, we eliminate network size effects by comparing the performance of the damaged network only with the corresponding original network.

We first consider the effects of targeted node removal on the passenger carrier networks.  Similarly to the analysis in \cite{holmePRE}, we target nodes iteratively by either degree or betweenness. That is, we remove the node with the highest degree or betweenness, then update each node's degree or betweenness and remove the node with the highest degree or betweenness.  Figure~\ref{figure:removal}(a) shows the size of the largest connected component, $S$, for iterative removal of the node with highest degree as a function of the proportion of nodes removed, $t$.  The inset of Fig.~\ref{figure:removal}(a) shows results for iterative removal of the node with highest betweenness.   The SW network stands out from the other passenger carriers, remaining fully connected after removing more than $30\%$ of nodes targeted by betweenness and more than $50\%$ of nodes targeted by degree.

The cost metric also reveals the resilience of SW.  Figure~\ref{figure:removal}(b) shows the normalized travel cost metric $\frac{T(M)}{T_0(M)}$ evaluated on the largest connected component $M$ of each passenger carrier's network as a function of the proportion of nodes removed by iterative degree-targeting, $t$.  Not only does the SW network stay fully connected after degree-targeted removal of a substantial fraction of nodes, but the remaining network continues to function nearly as efficiently as the undamaged network.  After removing the top $10\%$ of nodes, the total travel cost has only increased $4\%$ for SW while the cost of the next best carrier, $AA$, has increased by nearly $25\%$.  
Intuitively, a well-connected (high density) PP structure permits multiple nearly-shortest paths connecting most source-destination pairs.
In contrast, HS networks which route the majority of travel paths through relatively few (3-5) hubs perform worse under this metric since deletion of 
a nearby hub necessitates inefficient transcontinental crossings to the next-nearest hub in order to access the rest of the network.
Note, by the point $t=0.35$, $M$ for each HS network contains less than half of the nodes originally present. Due to the small remaining size, we can see $T/T_0$ dip for some networks. 

\begin{figure}[tbp]
\includegraphics[width=0.7\columnwidth, angle=270]{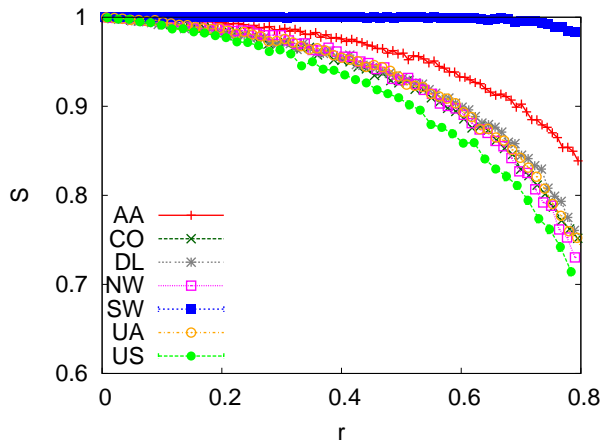}%
\caption{(Color online) Relative size of the largest connected component ($S$) of each passenger carrier's network as a function of the proportion of edges removed by random failure ($r$). Each data point is the  average over  $50$ independent realizations.   
Representative standard error is shown by the error bars on SW and US.}
\label{figure:edgedeletion}
\end{figure}

While using targeted removals is helpful for understanding worst-case scenarios, modeling random failures provides a different portrait of network resilience.  To this end, we consider the effects of random edge deletion.
Explicitly, we generate an ensemble of $50$ independent realizations (i.e., randomly selected sets of edges to delete) for each value of deleted edges considered.  Figure~\ref{figure:edgedeletion} shows the average value of $S$ (the relative size of the largest connected component) over the ensemble of 50 realizations as more edges are removed.   Remarkably, SW has nearly $ 98\%$ of its nodes in largest connected component even after the deletion of $80\%$ of its edges (and remains at 100\% connected for every realization in the ensemble until $30.8\%$ of the edges are removed).  In contrast, all of the other carriers have realizations that start losing full connectivity after the deletion of fewer than $2\%$ of edges, but note that the majority of the network remains connected. 
Thus the HS networks are fragile in the sense that even for low numbers of edges deleted,  a small set of nodes become completely disconnected from the network. This result is consistent with the prevalence of low-degree nodes occupying the low $k$-shells in the HS networks.  We also find that SW exhibits the slowest increase in the normalized travel cost metric under random edge deletion (not shown here), but this effect is much less pronounced than in Figure~\ref{figure:removal}(b).
 
We also find that all carriers are resilient to random node removal (not shown here). This does not come as a surprise, given that networks with  right skewed degree distributions are typically immune to random failures of their nodes. 
  
These results on resilience are consistent with our intuition that binary edge density alone is a strong predictor of network resilience.  SW is significantly more dense ($0.44$) than the HS airline with the next highest density, AA ($0.16$).  However, the detailed resilience portrait of a network depends not only on density but the specific wiring patterns present; two networks with the same density and radically different topologies (e.g. a star and a chain) could have very different resilience properties.  We will investigate this further in Sec.~\ref{sec:rewiring} by way of two different density-enhancing rewiring schemes.

\section{Constraint preserving rewirings}
It is of great interest to understand how to increase the resilience of an individual existing network.  We examine the effects of two rewiring schemes, called `Diamond' and `Chain,' which can increase binary edge density, and by consequence $k$-cores and resilience to node and edge deletion without increasing flight or airport requirements.  In order to boost the resilience of the an airline's route map, its unweighted binary network $G^c(N^c,E^c)$, we take advantage of the redundancy provided by its weighted network of actual flights, $W^c(N^c,E^c)$.  Each scheme involves rerouting flights within specific four-node motifs, found iteratively through search of each carrier's network, in a way that preserves both the number of flights and the in- and out-strength of each node.  We restrict our rewiring schemes to the undirected `Daily 1-flight minimum' weighted subnetwork for each carrier $c$ 
formed by rescaling all edge weights $s_{ij} \rightarrow \left\lfloor{\frac{s_{ij}}{365}}\right\rfloor$ and removing all edges with new weight less than $1$.  In cases where there is an asymmetric number of flights in each direction, we use the maximum as the undirected edge weight.  
 
\label{sec:rewiring}
\begin{figure}[htbp]
\includegraphics[width=\columnwidth, angle=0]{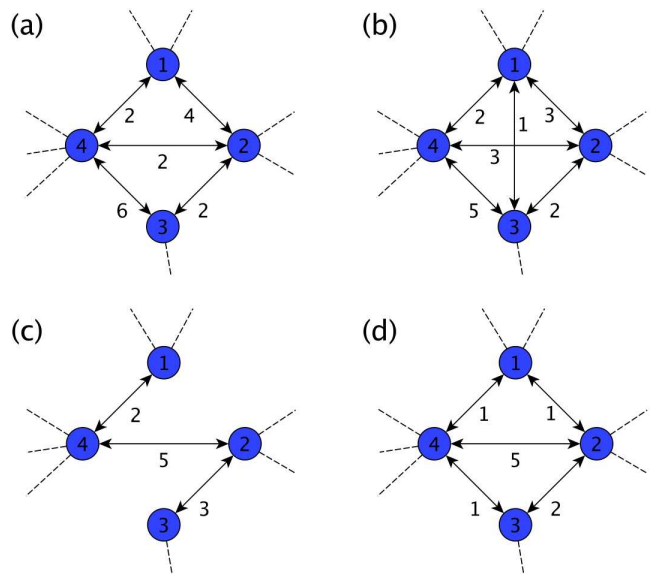}%
 \caption{(Color online)Two examples of strength-preserving rewirings which increase binary (unweighted) edge density and $k$-core of nodes.  In each, no explicit geography is implied by the layout and the edges between the nodes shown and the rest of the network are not shown.  (a) Diamond scheme: the initial logical weighted connectivity of a set of four nodes.    (b) Addition of a direct link between nodes $1$ and $3$ with adjustments of the weights on the existing links increases the coreness  of $1$ or $3$ or both.  The strength of each node  and the sum over all edge weights  remains constant despite the rewiring. (c) Chain scheme: the initial logical weighted connectivity of a set of four nodes.    (b) Addition of direct links between nodes $1$ and $2$ and nodes $3$ and $4$ with adjustments of the weights on the existing links increases the coreness  of $1$ or $3$ or both.  The strength of each node  and the sum over all edge weights  remains constant despite the rewiring.}
\label{fig:rewirings}
\end{figure}

In the `Diamond' scheme, we search for motifs with the structure shown in Fig.~\ref{fig:rewirings}(a), where the number of daily flights along the edge between $1$ and $2$ and the edge between $3$ and $4$ are at least $2$ (if there is only one flight between either pair we are not able to add a new binary edge and preserve gate requirements by shifting flights).  This motif is fairly common among hub-and-spoke networks, in which nodes $1$ and $3$ are spokes connected to hubs $2$ and $4$ but not to each other.  The missing connection to form a $4$-clique can be created by routing a small number of flights along the missing edge connecting nodes $1$ and $3$.  To preserve the gate requirements, a flight originally between nodes $3$ and $4$ is rerouted along $2$ and $4$ (see Fig.~\ref{fig:rewirings}(b)). In this manner the total number of flights (the sum over all edges) and the gate requirements (the in-strength and out-strength of each node) remain constant, while the addition of the edge connecting $1$ and $3$ raises the coreness of at least one of these two nodes.  To preferentially boost the most isolated nodes, we iteratively search for all such motifs and rewire the motif with the smallest sum of the degree of nodes $1$ and $3$; in the event that there are several qualifying motifs we select randomly among them.  While this rewiring scheme boosts edge density, it is restricted to boosting the resilience of nodes with degree at least $2$.  

On the other hand, the `Chain' scheme seeks to boost the resilience of the weakest nodes.  We search for motifs with the structure shown in Fig.~\ref{fig:rewirings}(c), where the number of daily flights along the edge between $1$ and $4$ and the edge between $2$ and $3$ are at least $2$.  Two additional binary connections are formed between $1$ and $2$ and $3$ and $4$ by transferring flights (see Fig.~\ref{fig:rewirings}(d)).  Such motifs are common in hub-and-spoke networks where spoke nodes ($1$ and $3$) lack connections to some of the hub nodes ($2$ and $4$).  Similarly to the Diamond scheme, we select motifs one at a time in which the sum of the degree of the weak nodes $1$ and $3$ is minimal, selecting randomly between qualifying motifs in the event that there is more than one.

We note the specific tradeoffs imposed by each rewiring.  As already mentioned, both the total number of flights and the gate requirements at each airport are preserved.  Additionally, under the Diamond rewiring, some of the demand between nodes $3$ and $4$, all of which was satisfied with direct flights before the rewiring, now must be satisfied by an indirect flight ($4$ to $2$ to $3$).  This inconvenience for some passengers is a tradeoff with respect to the convenience gained by other passengers who benefit from 
the new direct flight between nodes $1$ and $3$.  Similarly in the Chain scheme, some of the demand between nodes $1$ and $4$ or between $2$ and $3$, previously satisfied by direct flights may need to be satisfied by indirect flights (e.g. $1$ to $2$ to $4$).  In both rewiring schemes specific routes may be made more costly to passengers or the operator while others are made less costly, yet overall the airline gains flexibility through a denser route map.

We apply each of these rewiring schemes to the daily $1$-flight minimum network of each of the major passenger carriers, adding $10\%$ new edges, and examine the rewirings' effects on network resilience, measured according to the size of the largest connected component under degree-targeted node removal, and node betweenness.   

\begin{figure}[htbp]
\includegraphics[width=0.70\columnwidth, angle=270]{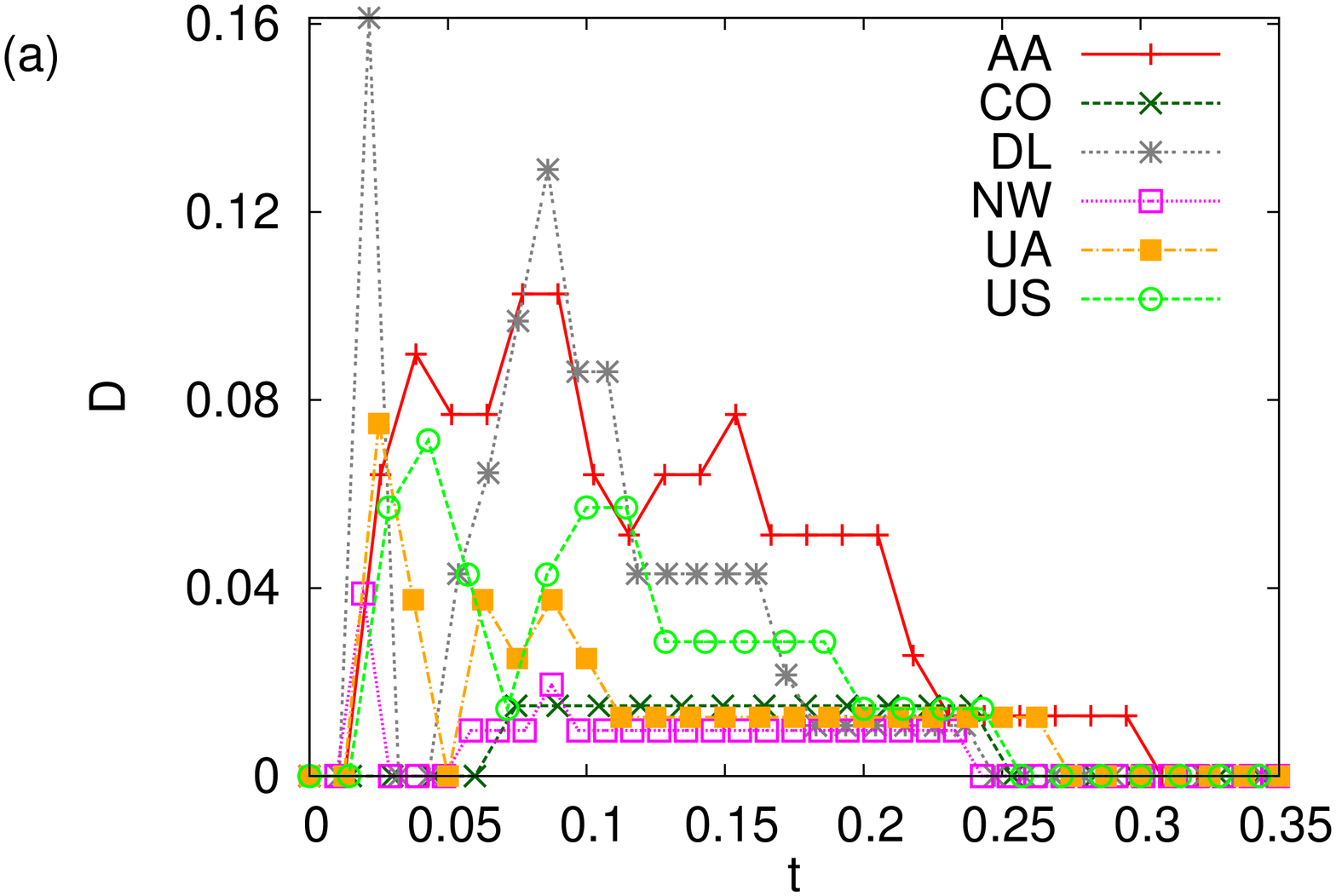}\hfill%
\includegraphics[width=0.70\columnwidth, angle=270]{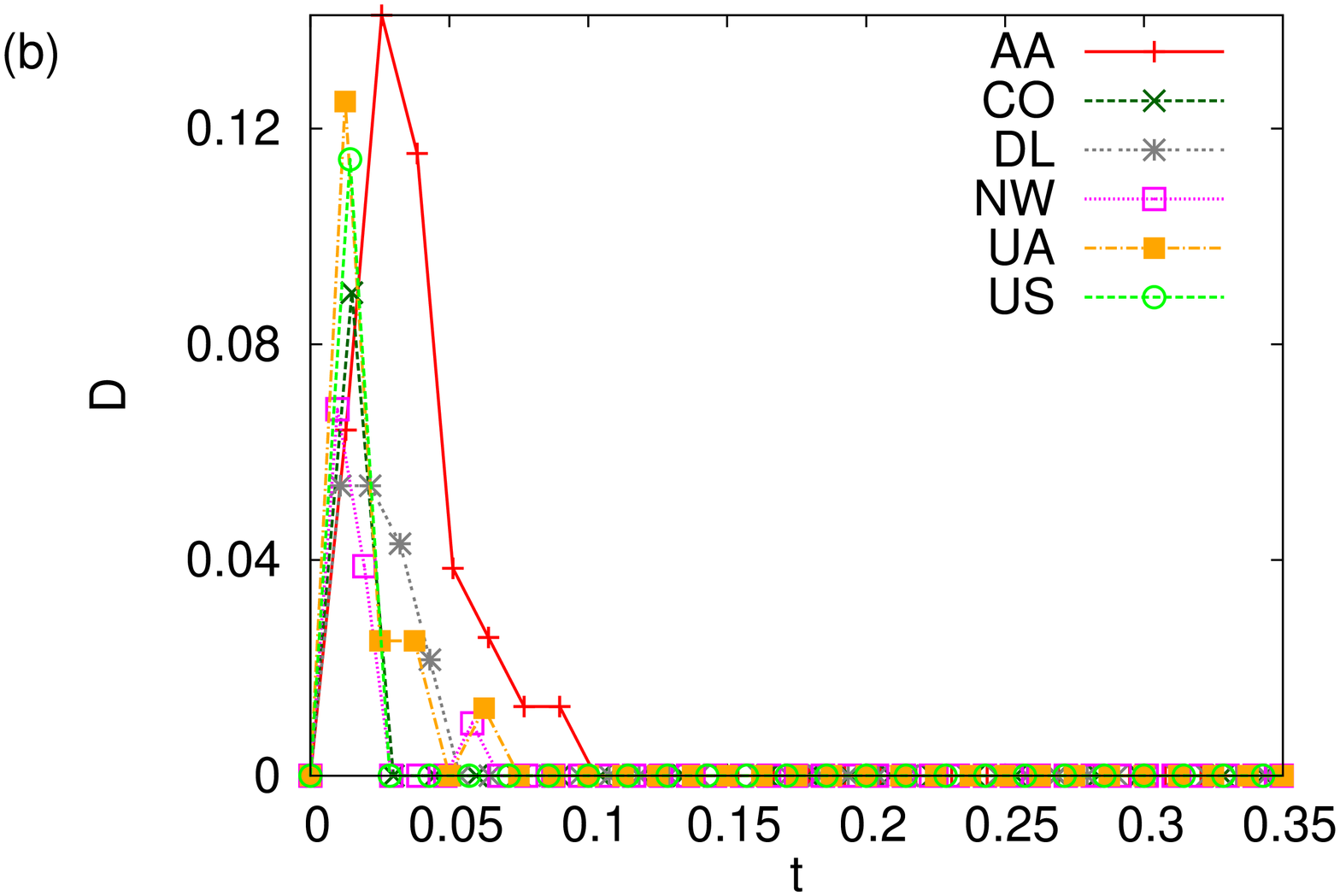}\hfill%
\includegraphics[width=0.70\columnwidth, angle=270]{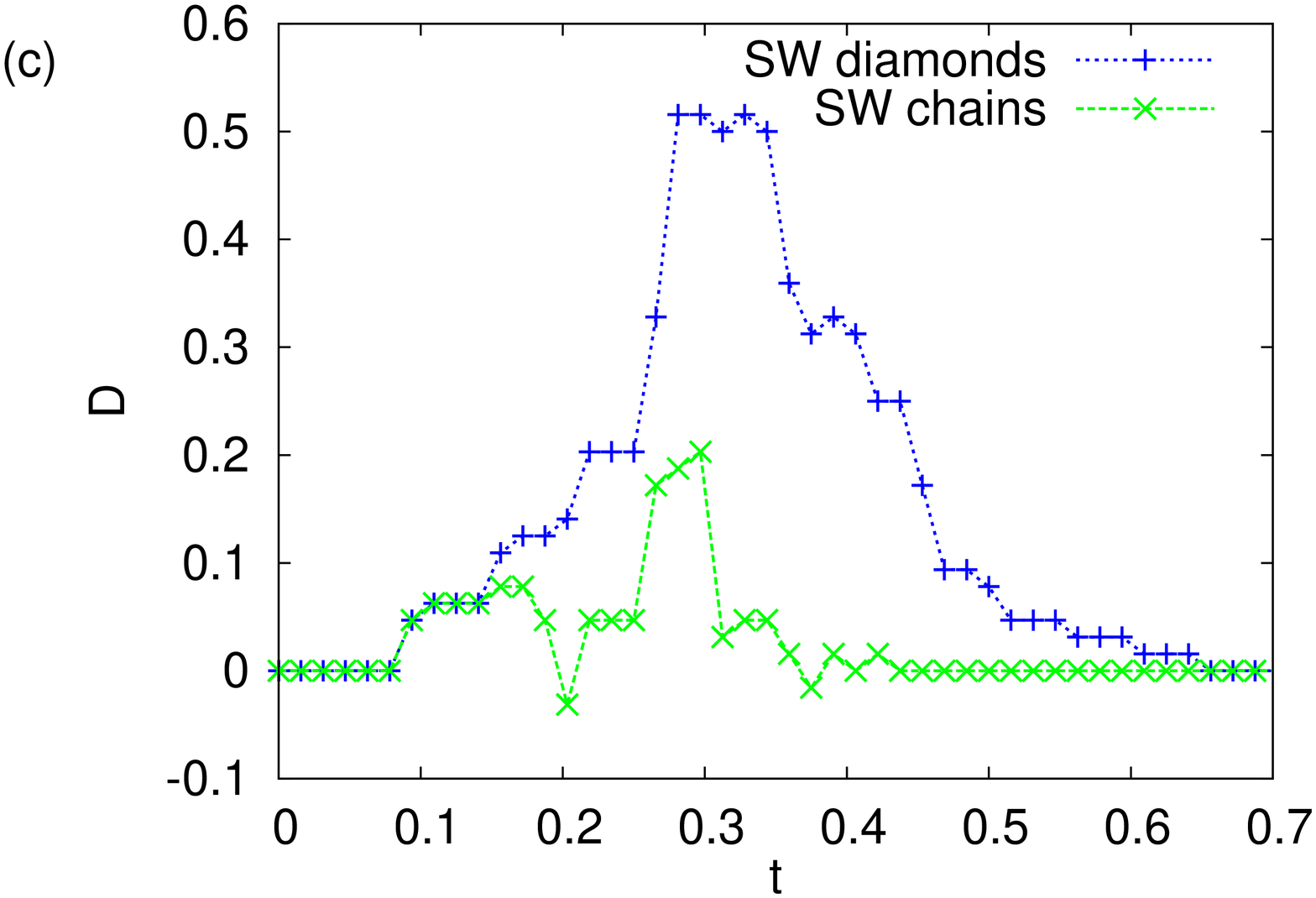}%
\caption{(Color online) The rewirings of Fig.~\ref{fig:rewirings} applied to the daily
$1$-flight minimum network of each carrier increases resilience.  Here $D=S(t)_{\mbox{rewired}} - S(t)_{\mbox{original}}$ with $S$ defined as in Fig.~\ref{figure:removal}.  Note that the `original' network we compare to is each carrier's daily $1$-flight minimum network.
(a) Diamond motif rewiring scheme applied to add $10\%$ new edges boosts resilience primarily to larger targeted disturbances.  (b) Chain motif rewiring scheme applied to add $10\%$ new edges boosts resilience to smaller targeted disturbances.  (c) Diamond and chain rewiring schemes applied to the SW network.  Note that gains in resilience occur in a later regime than other carriers since the original SW network remains well connected in the early regime.}\label{fig:rewired_resilience}
\end{figure}

Fig.\ref{fig:rewired_resilience} shows the effects of these two rewiring schemes on the resilience of the daily carrier networks.  In each, we plot $D$, the difference in the size of the connected component after removing a fraction of nodes by degree-targeted removal, $t$, between the rewired network and the original network, as a function of $t$.  As expected, the addition of edges via both schemes enhances the resilience of each network, though the resistance to different size disturbances depends on the scheme.  As seen in Fig.~\ref{fig:rewired_resilience}(a), while the Diamond scheme boosts the resilience of the networks to larger perturbations which knock out several of the most connected nodes, it offers no additional resilience to targeted perturbations which affect only the most connected node.  This is a consequence of the fact that this rewiring scheme can only be applied to nodes with degree at least $2$.  On the other hand, the Chain scheme can reinforce degree $1$ nodes and consequently boosts network resilience in the small perturbation regime, shown in Fig.~\ref{fig:rewired_resilience}(b).  The gain in resilience under the Chain scheme is most pronounced at the first nonzero data point in Fig.~\ref{fig:rewired_resilience}(b) (after the first node removal). Finally, the highly-connected structure of the SW network defers any gains in resilience until the larger perturbation regime. 

\begin{figure}[htbp]
\includegraphics[width=0.70\columnwidth, angle=270]{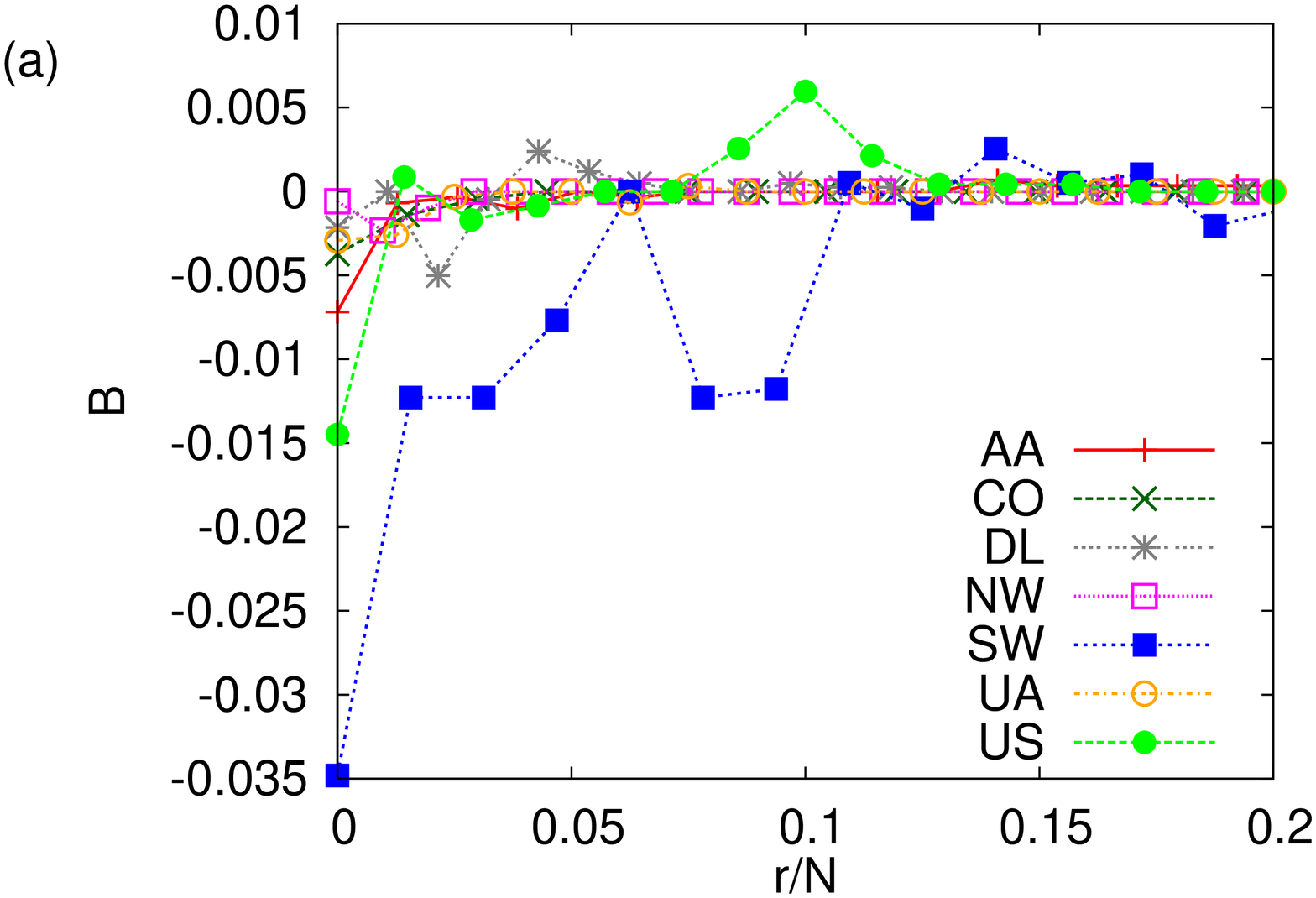}\hfill%
\includegraphics[width=0.70\columnwidth, angle=270]{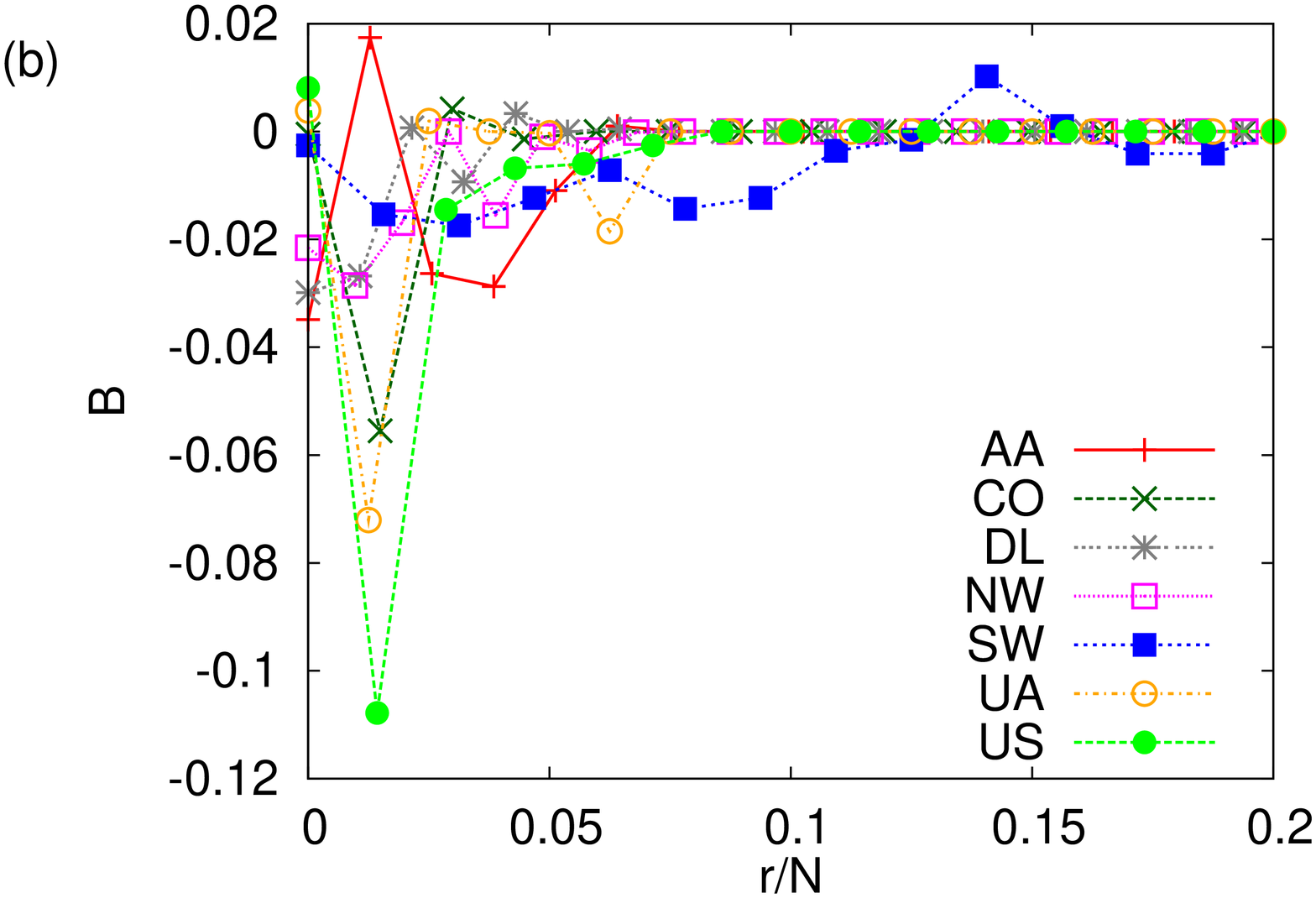}%
\caption{(Color online)  The rewirings of Fig.~\ref{fig:rewirings} applied to the daily
$1$-flight minimum network of each carrier modifies node betweenness.  Here we calculate the betweenness of each node using the geographic-distance weighted graph to determine shortest paths in both the original daily $1$-flight minimum network of each carrier and the rewired network.  We plot $B$, the difference in betweenness of the $r$-th highest betweenness-ranked node between the rewired network and the original network, against $r/N$ for each carrier.
(a) Diamond motif rewiring scheme applied to add $10\%$ new edges.  (b) Chain motif rewiring scheme applied to add $10\%$ new edges.}
\label{fig:rewired_btwn}
\end{figure}

Motivated by the fact that one consequence of the PP topology is that the shortest paths through the network can be distributed across many intermediate nodes rather than a few hubs, we examine the effects of the rewiring schemes on individual nodes' betweenness.  We calculate the betweenness of each node using the geographic-distance weighted graph to determine shortest paths in both the original daily $1$-flight minimum network of each carrier and the rewired network and plot $B$, the difference in betweenness of the $r$-th highest betweenness-ranked node between the rewired network and the original network, against $r/N$ for each carrier in~Fig.~\ref{fig:rewired_btwn}.  (Note that the rewiring scheme may actually shuffle the betweenness rank of some nodes).  While both schemes generally reduce the betweenness of the highest nodes, the Chain scheme has a more pronounced effect, particularly by reducing the betweenness of the top few hubs.  

It is noteworthy that while the two rewiring schemes increase edge density by the same amount, the specific resilience gains depend upon where these edges are added.  Furthermore, we emphasize that these rewiring schemes still respect the salient constraints of the original networks: the number of daily flights and the gate requirements at each airport.  While the specific many-variable optimization problems solved by the carriers may preclude such simple rewirings, this example suffices to show the existence of strength-preserving transformations which increase binary edge density and consequently network resilience to node and edge failure.

\section{Conclusion}
Using the abundant data available on the network structures of the major passenger airlines in the USA, we have we have studied the competing effects of efficiency and resilience in real-world networks.  
Although theoretical arguments suggest the asymptotic optimality of hub-and-spoke architectures for spatial transportation networks with transfer costs, we show that by including resilience into the considerations, in fact, point-to-point networks may be more desirable.  We have also shown that the degree assortativity coefficient of a network is sensitive to the existence of large hubs, and that structural analysis of networks in general should be augmented with other measures such as the Gini coefficient. Finally we explore the interplay between $k$-core structure and resilience of networks.  We introduce two different rewiring schemes which preserve node strength while boosting the coreness of either nodes with moderate $k$-cores or nodes with the lowest $k$-cores and show that the former boosts resilience to large perturbations while the latter boosts resilience to small perturbations.  Although developed in the context of the airline networks (where strength preservation is equivalent to preserving flight and gate requirements) the strength preserving rewiring schemes should be applicable to other networks in general.  Finally, although many other studies have found that airline networks show characteristics of power-law degree distributions~\cite{airChinaPRE04,airIndiaPhysica08}, we find that the degree distributions of the airline networks studied herein, including the aggregate views, are well described by simple exponential or cumulative log-normal distributions.

With regards to the airline networks specifically, we identify that of the seven largest USA passenger air carriers, Southwest Airlines has a remarkable topology especially with regards to its $k$-core structure, as more than half of all nodes belong to the $k_{\rm max}$-core.  We also establish the SW has extreme resilience to both random and targeted failures of nodes or edges.  We observe that the effect of targeted attack by betweenness, rather than by degree,  is significantly more pronounced on each carrier's network. This complements previous studies on the importance of network betweenness in general~\cite{holmePRE} and in airline networks in particular~\cite{GuimPNAS05}, underscoring that betweenness is an important criterion for consideration in critical infrastructure networks. 

Our findings raise the issue of whether hierarchical networks could be especially susceptible to targeted attacks or failures, given the rare population of the highest $k$-cores of such networks. 
The future design and operation of critical infrastructure may benefit from analyzing the tradeoffs of core versus peripheral placement of hub nodes, as mentioned in~\cite{doyle:2005pnas}. Hubs located in the core of a network substantially increase efficient connectivity yet are critical targets as without them, the network loses connectivity.  Hubs in the periphery (low $k$-cores) offer smaller benefits with respect to efficient connections, yet if they are disabled the connectivity of the core of the network remains largely unaffected.  Augmenting current studies on the optimal distribution of resources or facilities by including analysis of resilience properties of networks could increase their applicability.

\section{acknowledgement}

This work was supported in part by the Defense Threat Reduction Agency, Basic Research Award number HDTRA1-10-1-0088 and in part by the National Science Foundation through VIGRE Grant number DMS-0636297.

\end{document}